# Projected changes in synoptic circulations over Europe and their implications for summer precipitation: A CMIP6 perspective

Pedro Herrera-Lormendez[1], Amal John[2], Hervé Douville[2], Jörg Matschullat[1]

[1]Interdisciplinary Environmental Research Centre, TU Bergakademie Freiberg, Germany (Pedro.Herrera-Lormendez@ioez.tu-freiberg.de)

[2]Centre National de Recherches Météorologiques, Météo-France/CNRS, France.

**Abstract**

Projected changes in summer precipitation deficits partly depend on alterations in synoptic circulations. Here, the automated Jenkinson-Collison (JC) classification is used to assess the ability of twenty-one Global Climate Models (GCMs) to capture the frequency of recurring circulation types (CTs) and their implications for European daily precipitation intensities in summer (JJA). The ability of the GCMs to reproduce the observed present-day climate features is first evaluated. Most GCMs capture the observed links between mean directional flow characteristics of the CTs, and the occurrence of dry days and related dry months. The most robust relationships are found for anticyclonic and easterly CTs which are generally associated with higher than average occurrences of dry conditions.

Future changes in summer frequencies of the CTs are estimated in the high-emissions SSP5-8.5 scenario for the sake of a high signal-to-noise ratio. Our results reveal consistent changes, mainly in the zonal CTs. A robust decrease in frequency of the westerlies and increase in the frequency of easterly CTs favour more continental, dry and warm air masses over Central Europe. These dynamical changes are shown to enhance the projected summer drying over central and southern Europe.

# 1. Introduction

Synoptic surface circulations strongly influence the day-to-day weather of any region. Whether a place is dominated by a high- or low-pressure system, time of year, and most importantly, the source of the moving air mass (advective characteristics) drives their large-scale effects. Combining these large-scale patterns with local features can lead to weather extremes with devastating effects on society and economic activities. Furthermore, such patterns can influence longer-lasting events (e.g., heatwaves, droughts, floods) if some dry or wet circulation persists over anomalously long periods of time.

Several methodologies were developed to classify such patterns into categories known as Circulation Types (CTs). They all use atmospheric mean sea-level pressure data (MSLP), or 500 hPa geopotential heights (used for weather regimes), to group daily large-scale configurations (Huth et al. 2008). The Lamb Weather Types (Lamb 1972), one of the best-known classifications with its currently updated automated version, can be used to determine the synoptic patterns given a gridded MSLP dataset (Jenkinson and Collison 1977; Jones et al. 1993). This methodology has been widely used to investigate links between synoptic CTs and variables like temperature and precipitation; most prominently in the UK, where the classification was initially developed (Delaygue et al. 2019; Jones et al. 1993, 2016; Wilby et al. 1997) and later applied to continental Europe and other midlatitude regions (Brands 2022; Chen 2000; Demuzere et al. 2009; Donat et al. 2010; El Kenawy et al. 2014; Lhotka et al. 2020; Lorenzo 2011; Sarricolea Espinoza et al. 2014; Trigo 2000).

Atmospheric circulation is a major source of uncertainty in climate change projections, including in the midlatitudes (Shepherd et al. 2014; Oudar et al. 2020). Storylines based on illustrative circulation changes can be used to circumvent this obstacle (Shepherd et al., 2018), but do not lead to a comprehensive probabilistic assessment so that other strategies can be developed to account for circulation changes while assessing changes in regional climate.

The JC classification (JCC) often serves as a tool to better understand the influence of CTs on surface climate and, more recently, to investigate changes in seasonal circulation frequency under future climate change conditions. Larger datasets from reanalyses and growing outputs from Global Climate Models (GCMs) in the different generations of the Climate Model Intercomparison Project (CMIP) enabled this approach. However, applications are limited to the latitudes outside of the ± 22.5 degrees tropical belt (Fernández-Granja et al. 2022), given its dependency on the marked meridional pressure gradients that determine the structure and movement of these synoptic surface patterns. These applications largely suggest future weakening in meridional pressure gradients, directly triggering changes in the average synoptic conditions in Europe – more similar to conditions in the Mediterranean and the south of Europe after the mid-21$^{st}$ century (Herrera-Lormendez et al. 2021). This tendency also suggests seasonal changes in the mean zonal state of synoptic circulations like the westerlies and easterlies as well as cyclones directly influencing regional precipitation changes, e.g., over the British Isles (Burt and Ferranti 2012).

Some of these findings advocate for a strengthening of rainy westerly circulations over Northern Europe in winter and a corresponding decrease in easterlies and cyclonic circulations over the Mediterranean (Cahynová and Huth 2016; Demuzere et al. 2009; Hoy et al. 2014; Stryhal and Huth 2019b). On the other hand, summers show weakening westerly inflow with a corresponding increase in the frequency of some easterly advection over western and central Europe (Otero et al. 2018). Such projected increase of dry CTs to the detriment of rain-rich synoptic circulations supports current IPCC projections for wetter winters over Northern Europe and drier conditions southward, as well as a generalized decrease in summer precipitation across Europe and rainier conditions towards the northern part of the region (Douville et al. 2021; Santos et al. 2016).

Much previous work focuses on the winter season, given better discernability between the CTs due to stronger pressure gradients during this time of year, and the clearer signal these patterns have on surface climate variables. Nevertheless, summer will become a challenge for southern Europe, where precipitation amounts are expected to significantly decrease. Under a global warming scenario, where greenhouse gases continue to rise, the poleward shift of the Hadley cell edge will be enhanced (Grise and Davis 2020). Ramifications of these changes will influence, e.g., heat waves becoming more frequent, with higher mean duration, extent, and intensity (Schoetter et al. 2015) and more severe droughts with longer-lasting effects (Lhotka et al. 2020; Schwarzak et al. 2015). Nonetheless, some of the consequent changes in MSLP and associated circulation might directly strengthen or mitigate the projected precipitation changes over Europe (Belleflamme et al. 2015), as other large-scale phenomena, local thermodynamic factors and the dynamics within the changing CTs themselves might dominate over these alterations.

The objective of our study is to use a well-known and fixed classification of summer synoptic CTs to study changes in their relative frequencies and their influence on summer daily precipitation intensities over Europe. We investigate i) the links between European synoptic circulations and dry days; evaluate ii) how well GCMs reproduce the summer features of these patterns when compared to the ERA5 and E-OBS datasets; study iii) the influence of dry circulations on the sub-seasonal time scale during anomalously dry months; explore the iv) likely future frequency changes within the CTs towards the end of the current century and their impact on projected rainfall changes, and apply v) a linear decomposition of frequency and precipitation changes to better understand the contribution of CT frequency changes to projected precipitation anomalies.

## 2. Data and Method

### 2.1. Gridded JK classification

We follow the methodology proposed by Otero et al. (2018) by adapting the JCC with a moving central grid point to compute the corresponding gridded CTs over Europe from -15.5ºW to 30.5ºE and from 30.5 to 70.5ºN. This adapted version allows us to classify daily synoptic configurations at each grid cell of the selected region and then to assess the influence of the dominant CTs on any atmospheric variable during the chosen time frame. For instance, an anticyclone centred over Scandinavia will result in an easterly-dominant flow pattern over Germany and a southerly advection over the British Isles. The full classification initially considers 27 different synoptic patterns based on their dominant atmospheric pressure (cyclones and anticyclones) and their wind flow directions plus a "Low Flow" type. This last CT is characterised by days with very weak pressure gradients (persistent condition over the tropics), making it difficult to relate it to any dominant advection (see Otero et al., 2018, for more information on the method). Here, for the sake of simplicity, we aggregate some neighbouring CTs and only consider 11 types: the Anticyclonic (A) and Cyclonic (C) centre pressure systems, the corresponding eight prevailing directions of advection (NE, E, SE, S, etc.) and the remaining Low Flow (LF) type. Yet, our method could be easily applied to the original JCC and would lead to similar conclusions.

## 2.2. Influence of CTs on dry days and months

Daily MSLP from ERA5 reanalysis (Hersbach et al. 2020) was employed to calculate daily patterns and total rainfall from the E-OBS gridded dataset (Cornes et al. 2018) to investigate links between CTs and their influence on daily summer (JJA) precipitation patterns. ERA5 and E-OBS were used as reference datasets. A subset of 21 GCMS from CMIP6, based on the historical experiment (1951–2000) and the SSP5-8.5 scenario (2051–2100), were included in our analyses. All datasets were interpolated onto a 1°x1º common horizontal grid using bilinear and conservative methods to facilitate the model evaluation and intercomparison.

To investigate the CT influence on dry days, we compute the Conditional Probability of a Dry Day (CPDD) for each atmospheric pattern by accounting for all the summer days with total precipitation below one millimetre on each grid point of the domain, divided by the total summer relative frequency corresponding to the individual CTs.

To further explore their influence on dry months, we compute the Pearson 1-month Standardised Precipitation Index (SPI), given its advantage of using only precipitation as the parameter for its computation. The SPI calculation is based on the long-term precipitation record from the 1951-2000 period. This long-term record is fitted to a probability distribution, which is then transformed into a normal distribution so that the mean SPI for a desired location and period is zero (Edwards and McKee 1997). We use monthly aggregated rainfall from the different datasets resulting in a monthly-SPI timeseries that represents the degree of dryness (SPI < -1) and wetness (SPI > 1) over a chosen period. We employ 1-month SPI as it reflects better short-term dry conditions, which can affect, e.g., soil moisture and crop stress during summer (Svoboda et al. 2012). The SPI index was calculated using Adams's Climate Indices python package (2017). Dry or very dry months are kept in those months with SPI values below -1. The gridded monthly frequency anomalies of each CT, taking place during the given dry months, were computed to derive the monthly anomalous occurrences of these patterns during the summer season.

## 2.3. Attribution decomposition of future precipitation changes

Our main objective is to assess how changes in the CT frequency can contribute to an overall summer drying over Europe (Douville et al. 2021) and how model-dependent such a contribution can be. We tackle this by deriving future changes in the summer frequencies of the synoptic patterns, based on the difference between the late 21st century (2051–2100) and the historical reference period (1951–2000). We compute the spatial differences throughout the individual models and display the resulting multi-model ensemble (MME) mean and the corresponding multi-model agreement.

To investigate the attribution corresponding to the projected changes in CTs' frequencies on precipitation over Europe, we use the same linear decomposition equation as in Cattiaux et al. (2013) but using fixed rather than time-evolving circulation types. So doing, we can easily decompose future climatological changes in a variable X as follows:

$$\Delta^{F-P} = \overline{X}^F - \overline{X}^P = \underbrace{\sum_k \Delta f_k \cdot x_k^P}_{BC} + \underbrace{\sum_k f_k^P \cdot \Delta x_k}_{WC} + \underbrace{\sum_k \Delta f_k \cdot \Delta x_k}_{RES} \qquad \text{Eq. 1,}$$

where $f_k = \frac{N_k}{N}$ the frequency of occurrence of the $k$th circulation type and $x_k$ the precipitation conditional mean to regime $k$ defined by $x_k = \frac{1}{N_k} \sum_{i \in \Omega_k} X_{ik}$ with $\Omega_k$ the total ensemble of the $N_k$ days classified in the $k$th CT.

Equation 1 describes how changes in the frequency of the synoptic circulations contribute to changes in $\overline{X}$. This can be estimated considering the difference between two time periods, e.g., future (F) and present (P) climate. The first term is "Between-Class (BC)", representing the part of the total difference of precipitation due to frequency changes in the circulation types, i.e., an observed drying might be due to more frequent appearance of "dry" atmospheric circulations. The second term is "Within-Class (WC)", referring to the contribution of differences within the CTs which can be related to both thermodynamic and mesoscale dynamic processes, while the third one, "Residual (RES)", is defined as a mixing term. Here, the breakdown of total changes is applied at each grid cell over the whole European domain, allowing us to showcase the spatial distribution of the proposed decomposition.

## 3. Results

### 3.1. Spatial distribution and influence on dry days

Synoptic circulations owe their existence to the surface pressure gradients that shape the low-level horizontal winds and influence local weather through temperature and humidity advection. This influence becomes evident when visualizing the summer (JJA) spatial distribution of CPDD for various CTs (Figure 1a), computed over the 1951–2000 reference period using the ERA5 and EOBS datasets.

The JCC allows us to isolate synoptic conditions that feature a Low Flow (LF) pattern, without a strong advection. This CT is often observed in summer and is linked to a dominant ridge configuration (extension of a high-pressure system) which is distinct from the Anticyclonic CT. The LF pattern dominates over half of the summer days in the south of Europe, as expected from the weak meridional pressure gradients during JJA over this region.

The Anticyclonic circulation (A) is the most frequent pattern. Unlike the LF type, its influence is more evenly distributed over the continent. Its more significant predominance prevails over the continent and towards the Atlantic Ocean, given the extension of the Atlantic High that dominates summer months over the Azores. Contrastingly, the Cyclonic (C), the third dominant pattern, prevails over the northern latitudes (in the Northern Hemisphere) as these centres of low-pressure systems move incessantly from west to east. A similar pattern can be derived from the Westerlies (W), where they draw a well-defined belt along the midlatitudes that extends from the west of the British Isles to northern central Europe. The fifth most common pattern during summer is the Easterlies (E). Their occurrence, e.g., over central Europe, is often dictated by a dominant Anticyclone over Northern Europe; at times linked with heatwave events, dry days, and the advection of very warm air from the inner continent (Herrera-Lormendez et al. 2021).

Later, we investigate the probability of a dry-day occurrence during the occurrence of these circulations. Figure 1b shows the 1951–2000 summer (JJA) CPDD for the five main patterns. The LF synoptic type does not show a marked influence on precipitation over most of continental Europe. About half of the time, this pattern is linked with rainfall, albeit mainly over inland regions. We attribute this behaviour to dominant calm summer days in which warming over land can trigger small-scale convection events (summer storms) depending on moisture availability and previous weather conditions. Anticyclones confirm the well-known fact that they are responsible for dominant clear skies without rainfall. Exceptions to this rule appear in localised, orography-dominated areas such as the Alps, a possible result of mountain-favoured isolated convective storms. The opposite becomes clear for the Cyclonic type, where precipitation generally occurs when a cyclone moves through a region. Mediterranean coastal areas make for an exception, where this circulation takes place on less than 15% of the summer days. The westerlies stand out as one of the circulations that strongly modulate rainfall. They consistently favour its occurrence along the western coasts of Europe thanks to the transport of humid air from the Atlantic Ocean. The Easterlies mainly exert a dry influence over the continent. The western part of Europe experiences these patterns as dry; barely responsible for rainy events as they bring drier and warmer air from Central Europe. However, in some places like south-eastern Europe, they can produce precipitation given the influence of the Mediterranean and the Black Sea.

Some of the behaviours observed in the zonal circulations prevail in the remaining six CTs (Figure S1). The Easterly-like types with the lowest frequencies generally relate to the absence of precipitation over Western Europe. The rest of the Westerly-like types predominate, albeit less frequently, along the midlatitude belt, favouring precipitation throughout western Europe. It is also important to point out that when employing the ERA5 precipitation data to evaluate the CPDD, this dataset tends to consistently capture more rainy days (daily rainfall above 1 mm) compared to the E-OBS gridded dataset (Figure S7). This agrees with previous findings, pointing out that ERA5 tends to overestimate mean precipitation related to too many wet days over Europe compared to E-OBS (Bandhauer et al. 2022).

## 3.2. Model Evaluation

Our analysis is based on the latest generation of GCMs which have contributed to the sixth phase of the Coupled Model Intercomparison Project (Eyring et al., 2016). This new generation of GCMs showed an improved ability to reproduce the JCC-derived synoptic CTs compared to the former CMIP5 GCMs (Fernandez-Granja et al. 2021). Yet, models still show contrasted skills in reproducing the summer mean frequency of the classified CTs over Europe. To compare the frequencies obtained from ERA5 reanalysis and those from the models, we compute the MME mean of the 21 available GCMs over the reference period 1951–2000. Figure 2a shows the relative frequency biases for the five dominant CTs. The mean spatial relative frequency values are given in brackets for reference. By design of the method, GCMs are generally able to reproduce the spatial features of the various circulations. However, they underestimate the relative frequency of Low Flow patterns over the Mediterranean and of westerlies over the western midlatitude belt. Such discrepancies have been documented in previous evaluations (Huguenin et al. 2020; Stryhal and Huth 2019a, hinting to model biases that favour an excessively zonal flow across these locations. Cyclones and Easterlies are generally slightly overestimated in the MME over southern Europe. The differences in the anticyclones are less homogenous but generally overestimated throughout the western parts of Scandinavia.

Taylor diagrams in Figure 2b summarize the evaluation of summer relative frequencies of the patterns derived from the individual GCMs and the MME mean compared to those obtained from ERA5 reanalysis during the reference period. Correlation values are generally high in most CTs. This is also true for the LF, yet some discrepancies remain (STD < 1) between some models as they generally tend to capture fewer LF days. The A type portrays slightly better correlation values, yet their higher STD's reflect the more significant inter-model spread. The C and E types show better results in correlation coefficients and STD's. However, the Westerly type stands out as one of the CTs with the highest spread between the individual GCMs and the lowest correlation values. This behaviour is also observed in the remaining six CTs, where westerly-like CTs have the lowest performance among the climate models (Figure S2). These spatial biases are also found in the previous evaluation of the fifth generation of CMIP (Otero et al. 2018). However, smaller differences prevail in our findings as compared to CMIP5.

We also evaluate the models' ability to capture the probability of dry days during the occurrence of the different CTs. Figure 3a shows the MME differences relative to the E-OBS dataset of the summer CPDD (precipitation < 1 mm). Stippled areas show at least a 66% multi-model agreement in the sign of the biases and values in brackets state the spatial mean value of the CPDD. Overall, we find that GCMs tend to display more dry days over land areas. In contrast, more rainy days are captured over the northern parts of Europe and especially over Scandinavia during the five main circulations. Some of these differences are more evident within the Cyclonic types, where models generally identify more dry days southwards of the continent, whereas more rainy days take place overall in Scandinavia and the northern part of Britain. However, despite what might look like a strong signal in the south, the multi-model agreement in the sign of the biases remains confined to some areas only. These regions, where some of the models capture more dry days, coincide with those where these CT frequencies are generally slightly overestimated by the ensemble members. Anticyclones, however, show the smallest differences in capturing the lack of precipitation associated with the days dominated by these high-pressure systems. The Taylor diagrams (Figure 3b) shows how the MME and most of the models fail to properly capture the spatial distribution of these patterns (correlation coefficients below 0.7). Overall, correlation coefficient values are smaller in comparison to Fig 2b, due to higher spatial precipitation variability. The disagreement in the individual models is more visible in the W and C types with the lower STDs and correlation coefficients. However, various GCMs continue to differ significantly in their ability to accurately represent westerly patterns and their distribution of precipitation. This is strongly tied to the inability of some models to accurately predict the frequency of this CT at first, which results in a further decline in the CPDD's skill. These characteristics are prevalent in the westerly-related types shown in Figure S3.

### 3.3. Links with dry months

The day-to-day influence of CTs can influence the onset, duration, and strength of long-lasting events like dry spells and heatwaves. To address this, we investigated the fingerprint of the synoptic circulations on the monthly precipitation by computing the monthly SPI and analysing dry months with SPI < -1. This relationship derived from the E-OBS dataset (cf. Figure 4a) displays the average summer days per month without precipitation. Areas around the Mediterranean exhibit a large number of dry days where most of the summer months see no rain. Not surprisingly, areas with fewer dry days appear over western Scandinavia, the British Isles, central Europe, and the Alps. During dry months (cf. Figure 4b), the number of no-rainfall days increases homogeneously over most of the domain (except the Mediterranean) by about 4 to 6 days. Nevertheless, the individual influence of each CT on dry months varies significantly from one pattern to another. It has been previously shown that the increased monthly frequency of some anticyclonic atmospheric patterns can directly influence regional droughts (Phillips and McGregor 1998; Richardson et al. 2018). Therefore, it may be important to assess the long-term behaviour of these CTs in a warmer climate, as any significant change in their frequency may favour the occurrence of some extreme events (Neal et al. 2016).

To visualise this, we compute in each grid cell the monthly anomalous frequencies of the main synoptic circulations for dry versus all summer months. Figure 5a first show the present-day anomalous frequencies derived from the ERA5 reanalysis (derived CTs) and the E-OBS precipitation climatology (SPI values). Figure 5b shows those corresponding to the MME median estimated from the 21 CMIP6 GCMs. Stippling highlights the areas where at least 80% of the models agree with the sign of the ensemble mean anomalies. Overall, the models capture the sign and spatial distribution of the observed frequency anomalies although the limited length of the ERA5 record may introduce some noise at individual grid cells. In principle, the LF type plays a negligible role in the appearance of dry months as, generally, they do not show significant positive or negative anomalies in their occurrence. This comes as no surprise, since LF types favour mainly isolated precipitation events during summer. The Anticyclones and Easterly circulations considerably increase in their frequency during dry months. The increase in A type is more distributed over the domain, but with the highest model agreement (at least 80%) distinctly dominating over northern and eastern Europe. The Easterlies display an important westward increment in their monthly frequencies during dry summer months. Their influence is more marked along the Atlantic coast of Europe and over the British Islands. This anomalous occurrence of easterly advection can occur during Scandinavian blocking events that are often responsible for longer-lasting heatwaves and dry conditions as they favour the transport of dry, warm continental air from mainland Europe (Kautz et al. 2022; Pfahl 2014). North-Easterlies also appear considerably more often during dry months. Their seasonal frequency anomalies are very similar to those recognized in the E circulation. The southerly pattern shows similar influence but with corresponding smaller anomalies, given its scarcer dominance during the warm season (Figure S4). Contrary and as expected, the cyclonic and westerly circulations, responsible for many rainy days over Europe, considerably decrease in frequency during the dry months. This behaviour happens as the establishment of Anticyclones and Easterlies blocks the influence of humid westerly advection over the continent (Pfahl 2014).

## 3.4. Future changes and contribution of CT frequencies

Global models have shown to work reasonably well at reproducing the spatial characteristics of most of the synoptic patterns, albeit still with important biases in some of the individual members at replicating the Low Flow and westerly types. When reproducing precipitation within the different CTs, the models display generally more dry days over continental Europe during the occurrence of rainy types like the Cyclonic and Westerlies, whereas a general tendency prevails to capture more rainy days over Scandinavia when compared to E-OBS. Bearing in mind such limitations, our next step is to analyse changes in the CT frequencies in a warmer climate and their implication for dry days. For this purpose, we estimate the difference in summer frequencies and precipitation within the CTs between 2051-2100 in the SSP5-8.5 scenario and 1951-2000 in the historical simulation. For the sake of simplicity and equal treatment of each model, we only employ the first member (r1) of a subset of 21 GCMs from CMIP6. Figure 6 depicts these projected differences computed from the MME mean, with the hatched areas showing an 80% multi-model sign agreement. The dominant summer Low Flow type exhibits a generalized increase in frequency across the entire domain, which translates into more days with weather more akin to that of southern Europe and that is consistent with a projected weakening of near-surface wind speed over land under projected high greenhouse warmings (Deng et al. 2022). This increase in summer days dominated by weaker meridional pressure gradients has also been reported in previous studies focusing on central Europe (Herrera-Lormendez et al. 2021) and in the prior generation of CMIP GCMs (Otero et al. 2018). Anticyclones and Cyclones largely depict inverse responses. The North Atlantic storm track within the domain shows a consistent decrease in the frequency of cyclones in agreement with the projected near 5% decrease in the number of extratropical cyclones systems by the end of the 21$^{st}$ century (Priestley and Catto 2022). Contrastingly, the anticyclonic influence is expected to increase over the midlatitude belt extending from the British Isles to Scandinavia. As a response, the midlatitude summer cyclones that generally move eastwards will likely lose their influence over these regions. An apparent increase of cyclones as a response to a weaker influence of anticyclonic centres appears over the Mediterranean. However, areas where the model agreement is at least 80%, are limited to the western part of Spain for the signal in the cyclones. The Westerlies show a predominant future weakening that extends from the Mediterranean up to the British Isles. This decrease agrees with the likely increase in MSLP that would directly influence the appearance of this circulation moving further north during the warm season (Ozturk et al. 2021). In response, the Easterlies exhibit future strengthening over most of continental Europe. These patterns can advect dry and warm air from the inner continent during their occurrence and are rarely linked to precipitation over western Europe.

These patterns have a strong influence on day-to-day summer rainfall over Europe and on its spatial distribution. However, the question we address here is, do these frequency changes account for the total precipitation changes or are they also associated with significant precipitation changes conditional to these circulations? Figure 7d shows the future summer rainfall changes within each CT computed from the MME mean. The more considerable changes are found in the C and W types which are well-known strong modulators of summer precipitation over western Europe. Their projected regional frequency decrease over Europe is mirrored in their future precipitation influence, with negative anomalies dominating these areas. Interestingly, the frequency increase of easterlies over continental Europe does not show significant changes in rainfall over this region.

However, observed changes in precipitation over the continent are not limited to frequency changes among the different circulation types (BC effect). The changes within the CTs themselves (WC effect) can frequently dominate due to changing characteristics of associated weather within the patterns (Cahynová and Huth 2016; Küttel et al. 2011). To investigate whether the projected frequency changes are the primary reason for precipitation changes, we employ a linear decomposition model from Cattiaux et al. (2013), aiming to decompose the Between-Class (BC), Within-Class and Residual (RES) attributions of these changes (Equation 1). The resultant decomposition for the main five CTs is shown in Figure 7 in absolute seasonal precipitation changes in millimetres. The columns from left to right indicate the a) BC, b) WC and c) RES terms and d) the corresponding SSP5-8.5 projected precipitation changes within each CT resulting from the sum of the three terms (BC + WC + RES). The first noticeable result is the minimal contribution of the Residual term towards precipitation changes, which we will not be further discussed. Main contributions arise from both the BC and WC terms. However, the contributions do not always add to the projected change. In many of the cases these contributions can regionally offset or complement each other. This behaviour is very clear with the LF type, where BC suggests that an expected future frequency increase in LF conditions would lead to more precipitation, whereas the WC term would result in projected drying over southern Europe. The relevance of the WC term confirms that frequency changes are not primarily responsible for the observed drying over the Mediterranean-like climates (Seager et al. 2019). With Anticyclones, the WC term is the main contributor related to continental drying, whereas frequency changes (BC term) will be responsible for the observed wetter conditions over the northern domain. In both cases, the dominant areas with a frequency decrease of these circulations correspond to a precipitation reduction according to the BC term. However, for cyclones, despite the strong drying as a response to the frequency decrease along their northern typical path (shown in the BC term), the WC term drives the changes in the Mediterranean and northern Scandinavia.

Interestingly, for the Westerly types, the rainfall changes can be almost entirely attributed to the frequency changes expected due to their future weakening over Europe (i.e., drier future conditions) and strengthening along the northernmost latitudes (wetter future conditions). As for the Easterlies, the contributions arising from the BC and WC terms are partly cancelling each other. The future frequency decrease in Easterlies over continental Europe results, according to the BC term, in some of the summer precipitation to focus over the Alps and Balkans. The WC effect suggests a corresponding drying over southern Europe. The remaining CTs (Figure S6) follow the same behaviour as the main five types, in which the Westerly-like types have the BC term as the main contributor to precipitation changes, and the Easterly-like types equally produce opposite contributions from the BC and WC terms.

Cyclones and Westerly circulation types stand out as the most significant contributors to daily precipitation changes as they dominate the occurrence of rainfall in most regions over Europe. Yet, in most cases, the observed projected changes in summer precipitation cannot be fully explained by considering changes in CT frequencies only. The dominant sources for within-type variations can arise from a) subgrid-scale phenomena, like orographically forced rain, not being resolved by the datasets grid (Beck et al. 2007), b) modification in precipitation-generating processes and/or from model inability to accurately simulate these processes (Santos et al. 2016), c) strong land-ocean contrast in the response of near-surface relative humidity to global warming influencing the response of the water cycle (Byrne and O'Gorman 2016) or d) from SSTs' influence on driving changes in relative humidity on the regional scale (Douville et al. 2020).

## 4. Summary and conclusions

We have employed an adapted version of the JK classification at each grid cell of a 1° by 1° horizontal grid over Europe (-15 to 30ºE and 35 to 75ºN) to define CTs in summer from daily MSLP fields of the ERA5 reanalyses and of a selection of twenty-one GCMs from CMIP6. We have first evaluated the models' ability to reproduce the current frequencies of a reduced set of eleven CTs and to capture the spatial pattern of CPDD compared to the E-OBS gridded dataset. Our results demonstrate a strong relationship between CTs and dry days at all locations. Anticyclones and the Easterly types are the dominant synoptic patterns that favour the occurrence of dry days. In contrast, Cyclonic and Westerly circulations strongly favour precipitation over western Europe. Climate models are generally good at reproducing the spatial frequency characteristics of the synoptic types during the summer season. Biases remain in capturing Low Flow type frequencies over the Mediterranean. In comparison with

CMIP5 (Otero et al. 2018), models continue to underestimate the frequency of the Westerly types in the mid-latitude belt while overestimating Easterly types' during summer. Moreover, they tend to overestimate the occurrence of dry days during Cyclonic and Westerly CTs, while they generally overestimate the occurrence of wet days over Scandinavia in the majority of the CTs (except S and SE).

The extended influence of CTs on monthly precipitation was investigated. Summer months with SPI gridded values below -1 (i.e., dry months) have been distinguished. The anomalous frequency of the CTs during such dry months were evaluated. A clear regional relationship emerges between CTs responsible for dry days (A and E types) and their regional frequency increase during dry months. Overall, the MME approach as well as the individual models (at least 80% multi-model agreement over regions with the largest influences) accurately represent the spatial influence of CTs on dry months when compared to the reference E-OBS dataset. The more visible positive anomalies appear within the Anticyclonic and Easterly types. The Anticyclones have a more decisive influence over the Eastern part of Europe, whereas the Easterlies are more likely to increase in frequency during dry months along western Europe. In response, the rainy CTs (Cyclones and Westerlies) tend to appear less frequently.

Major changes in seasonal precipitation and dry days have been shown to arise partly from changes in the frequency of the zonal-like CTs, at least in the high-emission scenario SSP5-8.5 selected in our study. Days dominated by weak pressure gradients (Low Flow type) are projected to increase over most of the domain albeit more strongly over the Mediterranean. This increase will enhance the drying projected over southern Europe, likely due to WC contributions, for instance advection of air masses with lower relative humidity due to a land-sea warming contrast (Byrne et al. 2013), and to lower relative humidity, associated with a stronger inhibition of deep convection and more dry days (Douville and John, 2021). Part of the remaining projected drying over western and central Europe is strongly driven by weakening frequencies of the Westerly and Cyclonic types, and increasing frequencies of the Easterly and North-easterly types. Such changes are congruent with the projected extension of dry and stable summer conditions and the mean poleward shift of moisture corridors and associated atmospheric rivers over Europe (Sousa et al. 2020). However, it remains challenging to interpret the WC contributions as far as the vertical structure of the CTs in not analysed in more details. Further investigation is also needed to better understand the temperature and humidity features of the upstream air masses and their model-dependent sensitivity to climate change.

## Acknowledgements

We thankfully acknowledge funding through the EU International Training Network (ITN) Climate Advanced Forecasting of sub-seasonal Extremes (CAFE). The project is supported by the European Union's Horizon 2020 research and innovation programme under the Marie Skłodowska-Curie Grant Agreement No. 813844.

## References

Adams J, 2017. climate_indices. An open-source python library providing reference implementations of commonly-used climate indices. https://github.com/monocongo/climate_indices.

Bandhauer M, Isotta F, Lakatos M, Lussana C, Båserud L, Izsák B, Szentes O, Tveito OE, Frei C. 2022. Evaluation of daily precipitation analyses in E-OBS (v19.0e) and ERA5 by comparison to regional high-resolution datasets in European regions. *International Journal of Climatology*. John Wiley & Sons, Ltd, 42(2): 727–747. https://doi.org/10.1002/joc.7269.

Beck C, Jacobeit J, Jones PD. 2007. Frequency and within-type variations of large-scale circulation types and their effects on low-frequency climate variability in Central Europe since 1780. *International Journal of Climatology*, 27(4): 473–491. https://doi.org/10.1002/joc.1410.

Belleflamme A, Fettweis X, Erpicum M. 2015. Do global warming-induced circulation pattern changes affect temperature and precipitation over Europe during summer? *International Journal of Climatology*, 35(7): 1484–1499. https://doi.org/10.1002/joc.4070.

Brands S. 2022. A circulation-based performance atlas of the CMIP5 and 6 models for regional climate studies in the Northern Hemisphere mid-to-high latitudes. *Geoscientific Model Development*, 15(4): 1375–1411. https://doi.org/10.5194/gmd-15-1375-2022.

Burt TP, Ferranti EJS. 2012. Changing patterns of heavy rainfall in upland areas: a case study from northern England. *International Journal of Climatology*. John Wiley & Sons, Ltd, 32(4): 518–532. https://doi.org/10.1002/JOC.2287.

Byrne, M. P., & O'Gorman, P. A. 2013. Land–Ocean Warming Contrast over a Wide Range of Climates: Convective Quasi-Equilibrium Theory and Idealized Simulations. *Journal of Climate*, *26*(12), 4000–4016. https://doi.org/10.1175/JCLI-D-12-00262.1

Byrne MP, O'Gorman PA. 2016. Understanding Decreases in Land Relative Humidity with Global Warming: Conceptual Model and GCM Simulations. *Journal of Climate*. American Meteorological Society, 29(24): 9045–9061. https://doi.org/10.1175/JCLI-D-16-0351.1.

Cahynová M, Huth R. 2016. Atmospheric circulation influence on climatic trends in Europe: An analysis of circulation type classifications from the COST733 catalogue. *International Journal of Climatology*, 36(7): 2743–2760. https://doi.org/10.1002/joc.4003.

Cattiaux J, Douville H, Ribes A, Chauvin F, Plante C. 2013. Towards a better understanding of changes in wintertime cold extremes over Europe: A pilot study with CNRM and IPSL atmospheric models. *Climate Dynamics*. Springer Verlag, 40(9–10): 2433–2445. https://doi.org/10.1007/s00382-012-1436-7.

Chen D. 2000. A monthly circulation climatology for Sweden and its application to a winter temperature case study. *International Journal of Climatology*, 20(10): 1067–1076. https://doi.org/10.1002/1097-0088(200008)20:10<1067::AID-JOC528>3.0.CO;2-Q.

Cornes RC, van der Schrier G, van den Besselaar EJM, Jones PD. 2018. An Ensemble Version of the E-OBS Temperature and Precipitation Data Sets. *Journal of Geophysical Research: Atmospheres*. John Wiley & Sons, Ltd, 123(17): 9391–9409. https://doi.org/10.1029/2017JD028200.

Delaygue G, Brönnimann S, Jones PD, Blanchet J, Schwander M. 2019. Reconstruction of Lamb weather type series back to the eighteenth century. *Climate Dynamics*. Springer Verlag, 52(9–10): 6131–6148. https://doi.org/10.1007/s00382-018-4506-7.

Demuzere M, Werner M, van Lipzig NPM, Roeckner E. 2009. An analysis of present and future ECHAM5 pressure fields using a classification of circulation patterns. *International Journal of Climatology*. John Wiley & Sons, Ltd, 29(12): 1796–1810. https://doi.org/10.1002/joc.1821.

Deng, K., Liu, W., Azorin-Molina, C., Yang, S., Li, H., Zhang, G., Minola, L., & Chen, D. 2022. Terrestrial Stilling Projected to Continue in the Northern Hemisphere Mid-Latitudes. *Earth's Future*, *10*(7), e2021EF002448. https://doi.org/10.1029/2021EF002448.

Donat MG, Leckebusch GC, Pinto JG, Ulbrich U. 2010. European storminess and associated circulation weather types: Future changes deduced from a multi-model ensemble of GCM simulations. *Climate Research*, 42(1): 27–43. https://doi.org/10.3354/cr00853.

Douville H, Decharme B, Delire C, Colin J, Joetzjer E, Roehrig R, Saint-Martin D, Oudar T, Stchepounoff R, Voldoire A. 2020. Drivers of the enhanced decline of land near-surface relative humidity to abrupt 4xCO2 in CNRM-CM6-1. *Climate Dynamics*. Springer, 55(5–6): 1613–1629. https://doi.org/10.1007/s00382-020-05351-x.

Douville, H., & John, A. 2021. Fast adjustment versus slow SST-mediated response of daily precipitation statistics to abrupt 4xCO2. *Climate Dynamics*, *56*(3–4), 1083–1104. https://doi.org/10.1007/S00382-020-05522-W

Douville, H., K.Raghavan, J.Renwick, R.P.Allan, P.A.Arias, M.Barlow, R.Cerezo-Mota, A.Cherchi, T.Y.Gan, J.Gergis, D.Jiang, A. Khan, W. Pokam Mba, D. Rosenfeld, J. Tierney, and O. Zolina, 2021: Water Cycle Changes. In Climate Change 2021: The Physical Science Basis. Contribution of Working Group I to the Sixth Assessment Report of the Intergovernmental Panel on Climate Change [Masson-Delmotte, V., P. Zhai, A. Pirani, S.L. Connors, C. Péan, S. Berger, N. Caud, Y. Chen, L. Goldfarb, M.I. Gomis, M. Huang, K. Leitzell, E. Lonnoy, J.B.R. Matthews, T.K. Maycock, T. Waterfield, O. Yelekçi, R. Yu, and B. Zhou (eds.)]. Cambridge University Press, Cambridge, United Kingdom and New York, NY, USA, pp. 1055–1210, doi:10.1017/9781009157896.010.

Edwards, D. C., & McKee, T. B. 1997. Characteristics of 20th century drought in the United States at multiple time scales. *Climatology Report 97-2*, Department of Atmospheric Science, Colorado State University, Fort Collins, Colorado.

El Kenawy AM, McCabe MF, Stenchikov GL, Raj J. 2014. Multi-decadal classification of synoptic weather types, observed trends and links to rainfall characteristics over Saudi Arabia. *Frontiers in Environmental Science*. Frontiers Media S.A., 2(SEP): 37. https://doi.org/10.3389/fenvs.2014.00037.

Eyring, V., Bony, S., Meehl, G. A., Senior, C. A., Stevens, B., Stouffer, R. J., & Taylor, K. E. 2016. Overview of the Coupled Model Intercomparison Project Phase 6 (CMIP6) experimental design and organization. *Geoscientific Model Development*, *9*(5), 1937–1958. https://doi.org/10.5194/gmd-9-1937-2016.

Fernandez-Granja JA, Casanueva A, Bedia J, Fernandez J. 2021. Improved atmospheric circulation over Europe by the new generation of CMIP6 earth system models. *Climate Dynamics*. Springer Science and Business Media Deutschland GmbH, 1–14. https://doi.org/10.1007/s00382-021-05652-9.

Fernandez-Granja JA, Brands S, Casanueva A. 2022. Exploring the limits of the Jenkinson-Collison classification scheme for atmospheric circulation: A global assessment based on various reanalyses. . https://doi.org/10.21203/RS.3.RS-1415588/V1.

Grise KM, Davis SM. 2020. Hadley cell expansion in CMIP6 models. *Atmospheric Chemistry and Physics*. Copernicus GmbH, 20(9): 5249–5268. https://doi.org/10.5194/acp-20-5249-2020.

Herrera-Lormendez P, Mastrantonas N, Douville H, Hoy A, Matschullat J. 2021. Synoptic circulation changes over Central Europe from 1900 to 2100 – Reanalyses and CMIP6. *International Journal of Climatology*. John Wiley & Sons, Ltd. https://doi.org/10.1002/joc.7481.

Hersbach H, Bell B, Berrisford P, Hirahara S, Horányi A, Muñoz-Sabater J, Nicolas J, Peubey C, Radu R, Schepers D, Simmons A, Soci C, Abdalla S, Abellan X, Balsamo G, Bechtold P, Biavati G, Bidlot J, Bonavita M, De Chiara G, Dahlgren P, Dee D, Diamantakis M, Dragani R, Flemming J, Forbes R, Fuentes M, Geer

A, Haimberger L, Healy S, Hogan RJ, Hólm E, Janisková M, Keeley S, Laloyaux P, Lopez P, Lupu C, Radnoti G, de Rosnay P, Rozum I, Vamborg F, Villaume S, Thépaut JN. 2020. The ERA5 global reanalysis. *Quarterly Journal of the Royal Meteorological Society*. John Wiley and Sons Ltd, 146(730): 1999–2049. https://doi.org/10.1002/qj.3803.

Hoy A, Schucknecht A, Sepp M, Matschullat J. 2014. Large-scale synoptic types and their impact on European precipitation. *Theoretical and Applied Climatology*. Springer-Verlag Wien, 116(1–2): 19–35. https://doi.org/10.1007/s00704-013-0897-x.

Huguenin, M. F., Fischer, E. M., Kotlarski, S., Scherrer, S. C., Schwierz, C., & Knutti, R. 2020. Lack of Change in the Projected Frequency and Persistence of Atmospheric Circulation Types Over Central Europe. *Geophysical Research Letters*, *47*(9). https://doi.org/10.1029/2019GL086132.

Huth R, Beck C, Philipp A, Demuzere M, Ustrnul Z, Cahynová M, Kyselý J, Tveito OE. 2008. Classifications of Atmospheric Circulation Patterns. *Annals of the New York Academy of Sciences*, 1146(1): 105–152. https://doi.org/10.1196/annals.1446.019.

Jenkinson AF, Collison FP. 1977. *An Initial Climatology of Gales over the North Sea. Synoptic Climatology Branch Memorandum, No. 62., Meteorological Office, Bracknell*.

Jones PD, Hulme M, Briffa KR. 1993. A comparison of Lamb circulation types with an objective classification scheme. *International Journal of Climatology*. John Wiley & Sons, Ltd, 13(6): 655–663. https://doi.org/10.1002/joc.3370130606.

Jones PD, Harpham C, Lister D. 2016. Long-term trends in gale days and storminess for the Falkland Islands. *International Journal of Climatology*. John Wiley and Sons Ltd, 36(3): 1413–1427. https://doi.org/10.1002/joc.4434.

Kautz L-A, Martius O, Pfahl S, Pinto JG, Ramos AM, Sousa PM, Woollings T. 2022. Atmospheric blocking and weather extremes over the Euro-Atlantic sector – a review. *Weather and Climate Dynamics*, 3(1): 305–336. https://doi.org/10.5194/wcd-3-305-2022.

Küttel M, Luterbacher J, Wanner H. 2011. Multidecadal changes in winter circulation-climate relationship in Europe: Frequency variations, within-type modifications, and long-term trends. *Climate Dynamics*. Springer, 36(5–6): 957–972. https://doi.org/10.1007/s00382-009-0737-y.

Lamb HH. 1972. *British Isles weather types and a register of daily sequence of circulation patterns, 1861-1971: Geophysical Memoir*. HMSO.

Lhotka O, Trnka M, Kyselý J, Markonis Y, Balek J, Možný M. 2020. Atmospheric Circulation as a Factor Contributing to Increasing Drought Severity in Central Europe. *Journal of Geophysical Research: Atmospheres*. John Wiley & Sons, Ltd, 125(18): e2019JD032269. https://doi.org/10.1029/2019JD032269.

Lorenzo MN, Ramos AM, Taboada JJ, Gimeno L. 2011. Changes in present and future circulation types frequency in northwest Iberian Peninsula. *PLoS ONE*, 6(1): e16201. https://doi.org/10.1371/journal.pone.0016201.

Neal R, Fereday D, Crocker R, Comer RE. 2016. A flexible approach to defining weather patterns and their application in weather forecasting over Europe. *Meteorological Applications*. https://doi.org/10.1002/met.1563.

Otero N, Sillmann J, Butler T. 2018. Assessment of an extended version of the Jenkinson–Collison classification on CMIP5 models over Europe. *Climate Dynamics*. Springer Verlag, 50(5–6): 1559–1579. https://doi.org/10.1007/s00382-017-3705-y.

Oudar, T., Cattiaux, J., & Douville, H. 2020. Drivers of the Northern Extratropical Eddy-Driven Jet Change in CMIP5 and CMIP6 Models. *Geophysical Research Letters*, 47(8). https://doi.org/10.1029/2019GL086695

Ozturk T, Matte D, Christensen JH. 2021. Robustness of future atmospheric circulation changes over the EURO-CORDEX domain. *Climate Dynamics*. Springer Science and Business Media Deutschland GmbH, 1: 1–16. https://doi.org/10.1007/S00382-021-06069-0/FIGURES/17.

Pfahl S. 2014. Characterising the relationship between weather extremes in Europe and synoptic circulation features. *Natural Hazards and Earth System Sciences*. https://doi.org/10.5194/nhess-14-1461-2014.

Phillips ID, McGregor GR. 1998. The utility of a drought index for assessing the drought hazard in Devon and Cornwall, South West England. *Meteorological Applications*. Cambridge University Press, 5(4): 359–372. https://doi.org/10.1017/S1350482798000899.

Priestley, M. D. K., & Catto, J. L. 2022. Future changes in the extratropical storm tracks and cyclone intensity, wind speed, and structure. *Weather and Climate Dynamics*, 3(1), 337–360. https://doi.org/10.5194/wcd-3-337-2022.

Richardson D, Fowler HJ, Kilsby CG, Neal R. 2018. A new precipitation and drought climatology based on weather patterns. *International Journal of Climatology*. John Wiley and Sons Ltd, 38(2): 630–648. https://doi.org/10.1002/joc.5199.

Santos JA, Belo-Pereira M, Fraga H, Pinto JG. 2016. Understanding climate change projections for precipitation over western europe with a weather typing approach. *Journal of Geophysical Research*. Wiley-Blackwell, 121(3): 1170–1189. https://doi.org/10.1002/2015JD024399.

Sarricolea Espinoza P, Meseguer-Ruiz Ó, Martín-Vide J, Martín-Vide J. 2014. Variabilidad y tendencias climáticas en Chile central en el período 1950-2010 mediante la determinación de los tipos sinópticos de Jenkinson y Collison. *Boletín de la Asociación de Geógrafos Españoles*, 64: 227–247. https://doi.org/https://doi.org/10.21138/bage.1688.

Schoetter R, Cattiaux J, Douville H. 2015. Changes of western European heat wave characteristics projected by the CMIP5 ensemble. *Climate Dynamics*. Springer Verlag, 45(5–6): 1601–1616. https://doi.org/10.1007/s00382-014-2434-8.

Schwarzak S, Hänsel S, Matschullat J. 2015. Projected changes in extreme precipitation characteristics for Central Eastern Germany (21st century, model-based analysis). *International Journal of Climatology*. John Wiley and Sons Ltd, 35(10): 2724–2734. https://doi.org/10.1002/joc.4166.

Seager R, Osborn TJ, Kushnir Y, Simpson IR, Nakamura J, Liu H. 2019. Climate Variability and Change of Mediterranean-Type Climates. *Journal of Climate*. American Meteorological Society, 32(10): 2887–2915. https://doi.org/10.1175/JCLI-D-18-0472.1.

Shepherd, T. G. 2014. Atmospheric circulation as a source of uncertainty in climate change projections. In *Nature Geoscience* (Vol. 7, Issue 10, pp. 703–708). Nature Publishing Group. https://doi.org/10.1038/NGEO2253.

Shepherd, T. G., Boyd, E., Calel, R. A., Chapman, S. C., Dessai, S., Dima-West, I. M., Fowler, H. J., James, R., Maraun, D., Martius, O., Senior, C. A., Sobel, A. H., Stainforth, D. A., Tett, S. F. B., Trenberth, K. E., van den Hurk, B. J. J. M., Watkins, N. W., Wilby, R. L., & Zenghelis, D. A. 2018. Storylines: an alternative approach to representing uncertainty in physical aspects of climate change. *Climatic Change*, *151*(3–4), 555–571. https://doi.org/10.1007/S10584-018-2317-9

Sousa PM, Ramos AM, Raible CC, Messmer M, Tomé R, Pinto JG, Trigo RM. 2020. North Atlantic Integrated Water Vapor Transport—From 850 to 2100 CE: Impacts on Western European Rainfall. *Journal of Climate*. American Meteorological Society, 33(1): 263–279. https://doi.org/10.1175/JCLI-D-19-0348.1.

Stryhal, J., & Huth, R. 2019a. Classifications of winter atmospheric circulation patterns: validation of CMIP5 GCMs over Europe and the North Atlantic. *Climate Dynamics*, *52*(5–6), 3575–3598. https://doi.org/10.1007/s00382-018-4344-7.

Stryhal J, Huth R. 2019b. Trends in winter circulation over the British Isles and central Europe in twenty-first century projections by 25 CMIP5 GCMs. *Climate Dynamics*. Springer, 52(1–2): 1063–1075. https://doi.org/10.1007/s00382-018-4178-3.

Svoboda M, Hayes M, Wood DA, World Meteorological Organization (WMO), 2012. Standardized Precipitation Index User Guide. WMO-No. 1090. WMO. Geneva: WMO. http://library.wmo.int/opac/index.php?lvl=notice_display&id=13682

Trigo RM, DaCamara CC. 2000. Circulation weather types and their influence on the precipitation regime in Portugal. *International Journal of Climatology*. John Wiley & Sons, Ltd, 20(13): 1559–1581. https://doi.org/10.1002/1097-0088(20001115)20:13<1559::AID-JOC555>3.0.CO;2-5.

Wilby RL, O'Hare G, Barnsley N. 1997. The North Atlantic Oscillation and British Isles climate variability, 1865–1996. *Weather*. John Wiley & Sons, Ltd, 52(9): 266–276. https://doi.org/10.1002/j.1477-8696.1997.tb06323.x.

# Figures

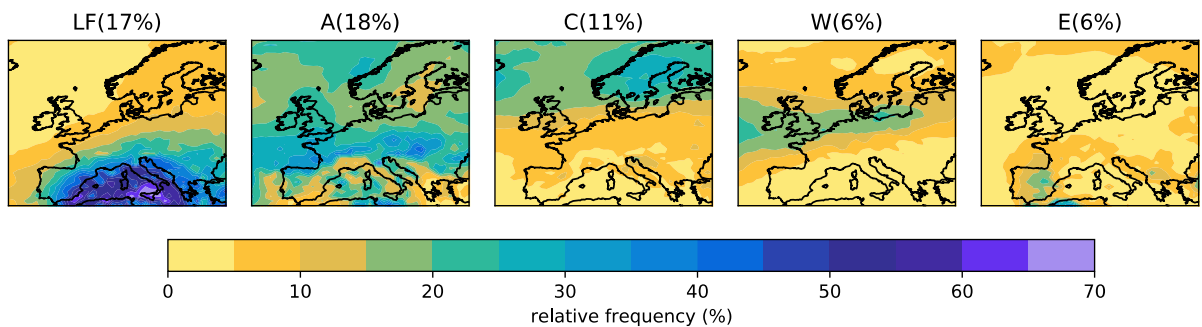

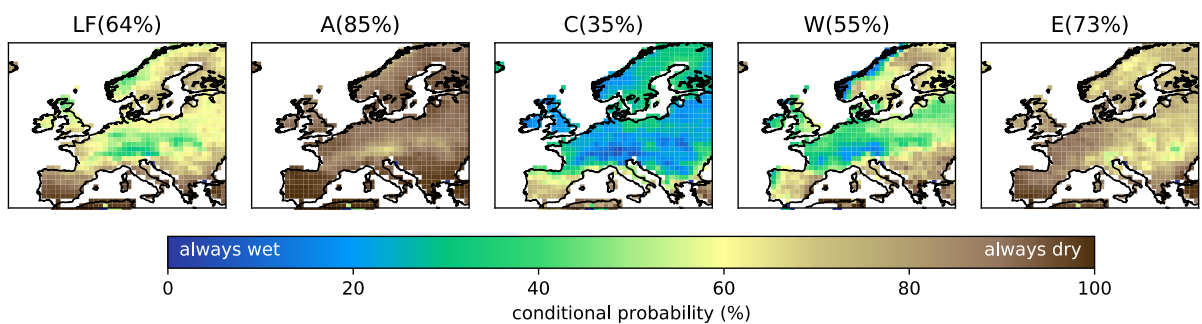

**Figure 1.** Summer (JJA) relative frequencies of the main five CTs derived from ERA5 (a) and CPDD corresponding to E-OBS (b) for the 1951–2000 period. Corresponding spatial mean values are shown in brackets.

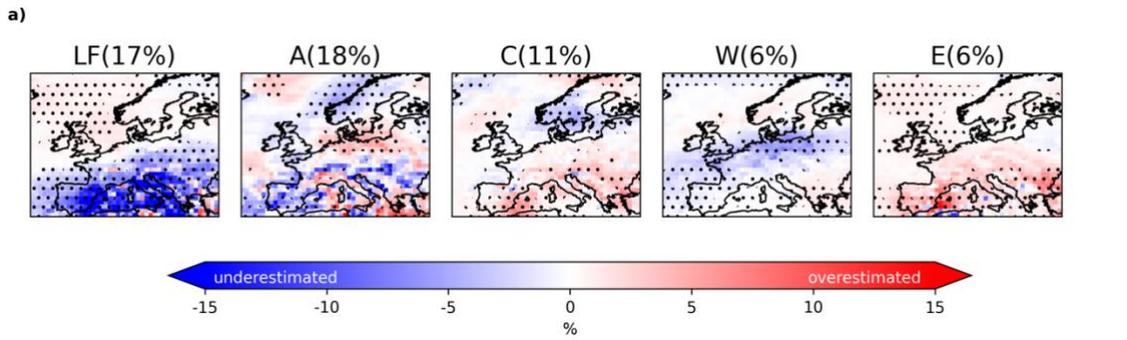
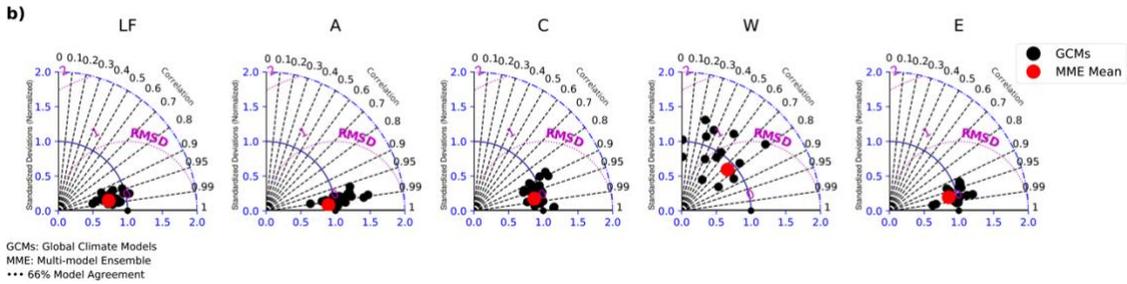

**Figure 2.** Summer (JJA) differences in MME Mean minus ERA5 relative frequencies (a) and corresponding Taylor diagrams (b) for the five main CTs. Stippling in "a" indicates a multi-model sign agreement of at least 66.6%. The mean spatial JJA relative frequency of each CT is indicated in brackets for the 1951-2000 period.

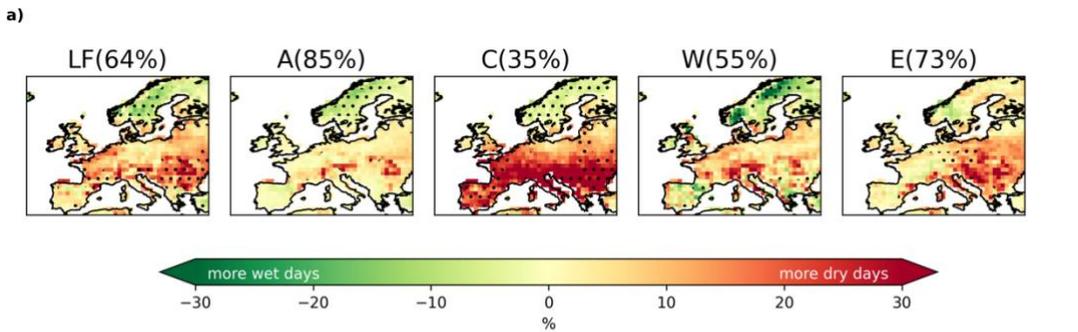
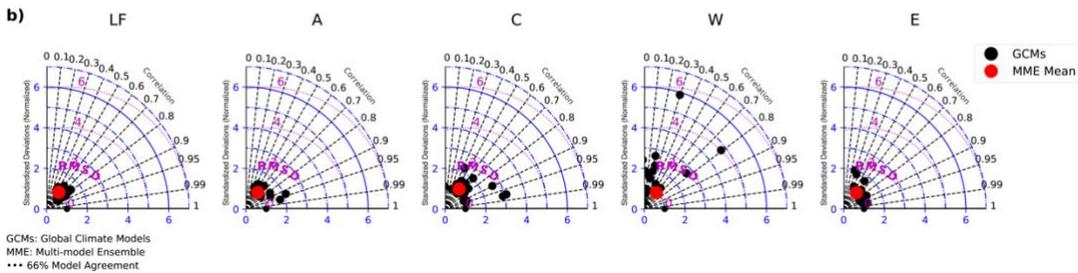

**Figure 3.** Summer (JJA) differences in MME Mean vs E-OBS CPDD (a) and corresponding Taylor diagrams (b) for the five main CTs. Stippling in "a" indicates a multi-model sign agreement of at least 66.6%. The mean spatial JJA CPDD of each CT is indicated in brackets for the 1951-2000 period.

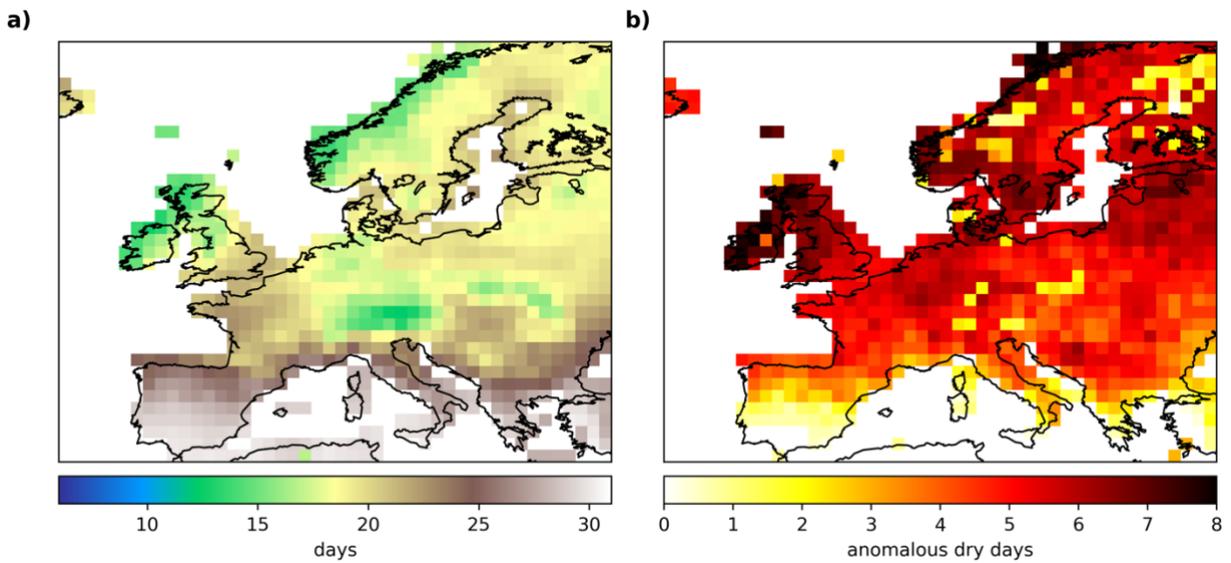

**Figure 4.** Summer (JJA) average number of dry days per month (a) and anomalous number of dry days during dry months with SPI gridded values below -1 (b) for the corresponding 1951-2000 period derived from the E-OBS dataset.

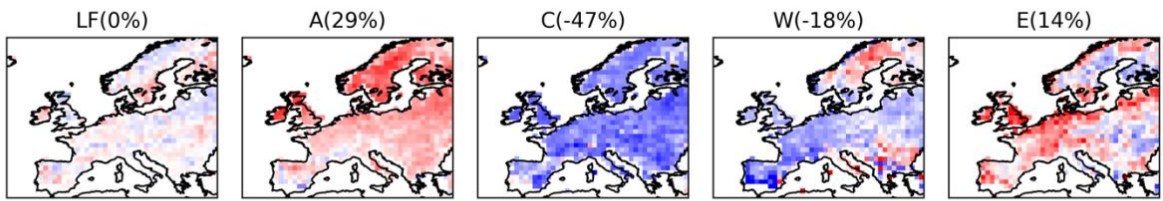

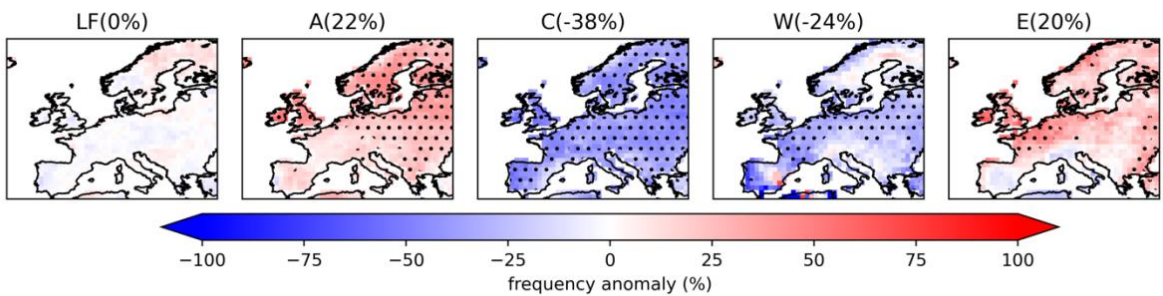

**Figure 5.** Summer (JJA) monthly frequency anomalies of the five main CTs in ERA5 and E-OBS datasets (a) and the CMIP6 MME Median (b) during months with SPI values below minus 1 corresponding to the 1951-2000 period. Stippling in "b" indicates a multi-model sign agreement of at least 80% of GCMs. Mean spatial frequency anomalies of the five CTs are shown in brackets.

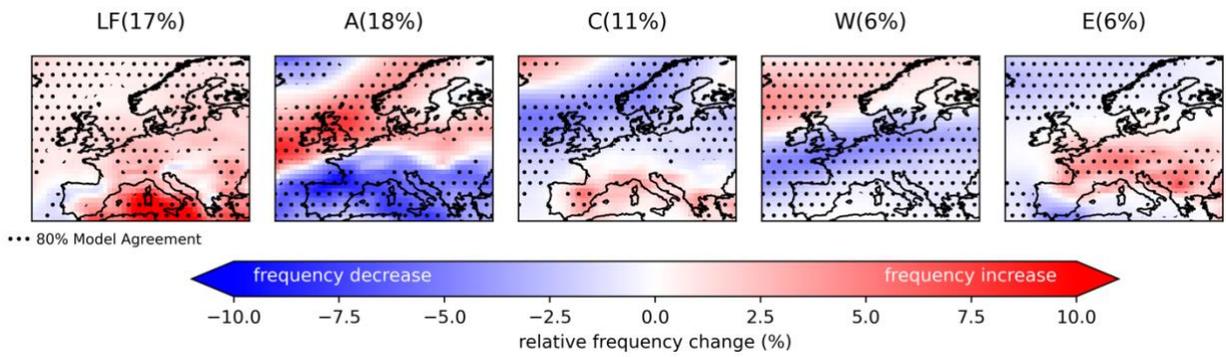

**Figure 6.** Summer (JJA) future changes in relative frequencies of five main circulation types over Europe. Stippling indicates 80% multi-model sign agreement and the mean seasonal relative frequencies for the reference period are shown in brackets. The changes are computed based on the future (2051-2100) minus present (1951-2000) periods.

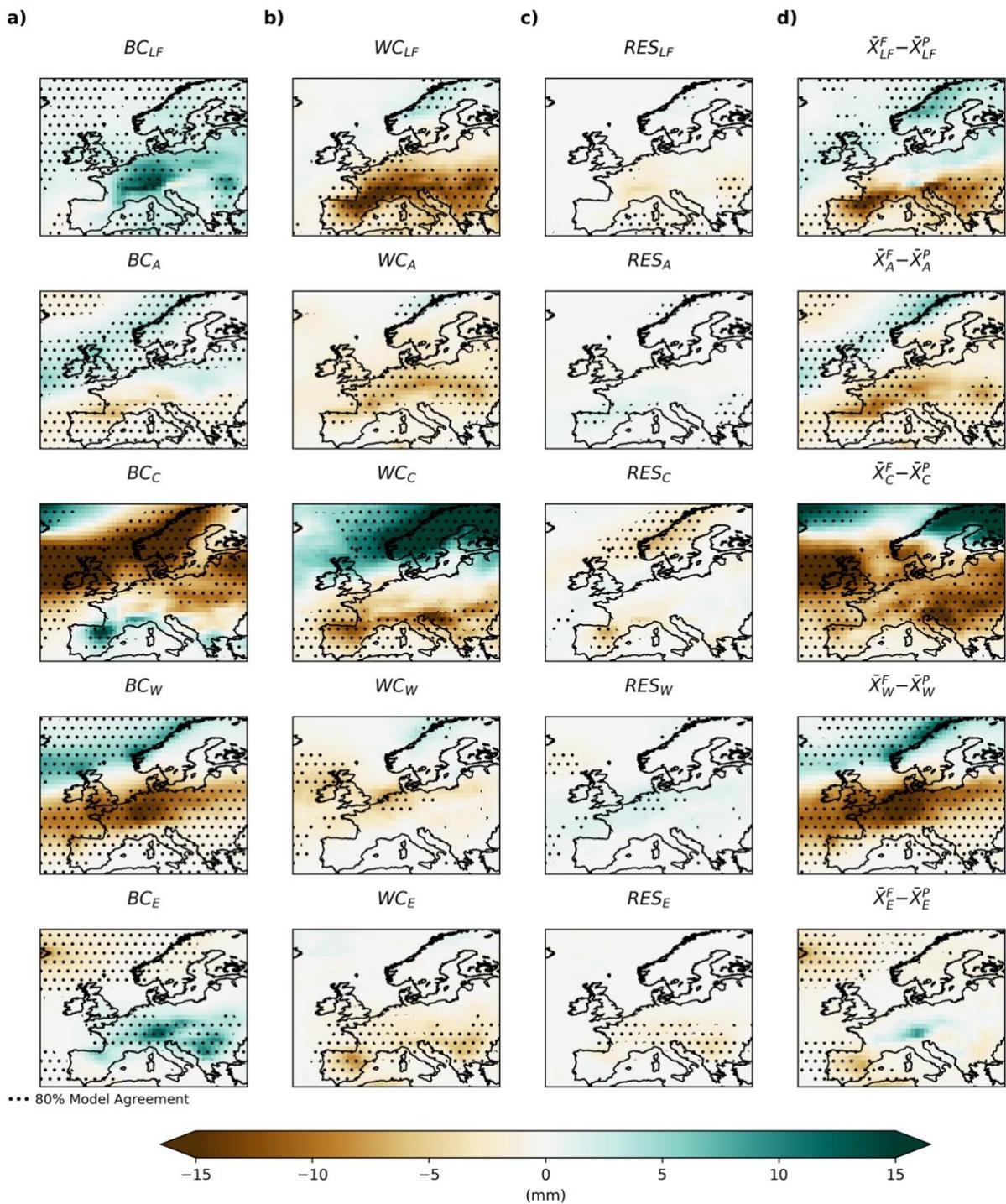

**Figure 7.** Decomposition of the summer (JJA) MME- Future(F) - Historical(P) difference in seasonal precipitation given equation 1. Stippling indicates 80% sign multi-model agreement. Columns show the a) BC, b) WC, c) RES terms and the d) absolute future precipitation change ($\overline{X}^F - \overline{X}^P$), while rows show the five main circulation types (LF, A, C, W and E).